\documentclass[prl,twocolumn,showpacs,amsmath,superscriptaddress,amssymb]{revtex4}

\usepackage{graphicx}
\usepackage{dcolumn}
\usepackage{bm}

\begin{document}


\title{Berry curvature in magnon-phonon hybrid systems}
\author{Ryuji Takahashi}
\email{ryuji.takahashi@riken.jp}
\affiliation{Department of Applied Physics, The University of Tokyo, 7-3-1, 
Hongo, Bunkyo-ku, Tokyo 113-8656, Japan}

\affiliation{Condensed Matter Theory Laboratory, RIKEN, Wako, Saitama, Japan}

\author{Naoto Nagaosa}
\affiliation{Department of Applied Physics, The University of Tokyo, 7-3-1, 
Hongo, Bunkyo-ku, Tokyo 113-8656, Japan}
\affiliation{RIKEN Center for Emergent Matter Science (CEMS), 
Wako, Saitama 351-0198, Japan}

\date{\today}

\begin{abstract}
We theoretically study the Berry curvature 
of the magnon induced by the hybridization with the acoustic phonons via the spin-orbit and
dipolar interactions. 
We first discuss the magnon-phonon hybridization via the dipolar interaction, and show that the dispersions have gapless points in momentum space, 
some of which form a loop.
Next, when both spin-orbit and dipolar interactions are considered, we show anisotropic texture of the Berry curvature and its divergence with and without gap-closing.
Realistic evaluation of the consequent 
anomalous velocity is given for yttrium iron garnet.
\end{abstract} 

\maketitle

{\it Introduction.--} 
The intensive studies on anomalous Hall effect in ferromagnets have revealed several microscopic mechanisms~\cite{NagaosaN10}. Particularly, the intrinsic mechanism is neatly described by Berry phase/curvature~\cite{BerryMV84}. The wave-packet formalism offers an intuitive picture on this mechanism~\cite{ XiaoD10}.   
The Berry phase/curvature acts as the emergent vector potential/magnetic field due to the constraint of the Hilbert space~\cite{NagaosaN12},
 and therefore it is also relevant to magnons, which obey Bose-Einstein statistics, as recent studies of the magnon Hall effect have shown~\cite{ KatsuraH10, OnoseY10, MatsumotoR11,ShindouR13}.


Recently, the dynamics of magnons has been studied by the advanced optical
spectroscopic method, in which propagating magnons have been observed~\cite{DemokritovSO01,DemokritovSO08}.
Since the magnon conveys information of the spin, it is important to investigate its dynamics. 
Creation, manipulations, and detection of magnons are called magnonics, and a variety of applications are expected in this emerging field~\cite{Kruglyak10}.
Meanwhile, it is well known that the magnon interacts with phonons~\cite{Kittel58}, and therefore phonons play an important role in magnonics, e.g., the emergence of the bottleneck effect of the magnon condensation at the hybridization region of magnon-phonon dispersion~\cite{HillebrandsAR12}. In addition, it has been demonstrated that the coherent phonon excitation can create the magnons through their hybridization~\cite{OgawaN15}. 
The excited quasiparticle has a finite width in both position and momentum spaces in general, and therefore, the motion of the wavepacket is relevant in the magnon-phonon hybrid systems.
 

In this paper, we theoretically study the coupling between magnons and acoustic phonons through the dipole-dipole interaction (DDI) in a simple model without sublattice structure, and their influence on the wavepacket dynamics of the magnon is investigated. 
 Historically, the magnon-phonon interaction by the magnetoelasticity, i.e., the spin-orbit interaction (SOI), has been studied for a long time~\cite{Kittel58, SchlomannE60,Gurevich96, SergaAA10,RueckriegelA14, AgrawalM12}. 
We find that the magnon-phonon hybridization by the DDI induces the anti-crossing as with the SOI and hence the Berry curvature, i.e., anomalous velocity, of the magnon-phonon hybrid.
Note that the phonon-magnon interaction is ubiquitous even for a simple ferromagnets without SOI or the sublattice structures, due to the presence of the DDI. 
We calculate the Berry curvature for the longitudinal acoustic (LA) and transverse acoustic (TA) modes, respectively, and find the high anisotropy.
 In the case of the LA mode, the divergence of the Berry curvature occurs at which the hybridization disappears, and some of these points form a loop.
For the TA modes, despite of the absence of gapless points, the divergence of the Berry curvature occurs.
We also find the rapid sign changes of the Berry curvature in the anti-crossing region.
Finally, we give a scheme for observing the Hall current of the magnon.

The spin Hamiltonian of a ferromagnet is 
\begin{eqnarray}
H_{m}&=&-\mu h\sum_{l}S_{l}^{z}-2J\sum_{\langle l,m \rangle}\mathbf{S}_{l}\cdot\mathbf{S}_{m}\nonumber\\
&+&\sum_{\l\not= m}\frac{\mu^2 [\mathbf{r}_{lm}^2\mathbf{S}_{l}\cdot\mathbf{S}_{m}-3(\mathbf{r}_{lm}\cdot\mathbf{S}_{l})(\mathbf{r}_{lm}\cdot\mathbf{S}_{m})]}{2|\mathbf{r}_{lm}|^5},
\label{eq:hamiltonianm}
\end{eqnarray}
where $\mathbf{S}_{m}$ is the spin at site $m$, 
and $\mathbf{r}_{lm}=\mathbf{r}_{l}-\mathbf{r}_{m}$ is the vector from site $m$ to $l$.
 $\mu=g\mu_{B}$ is the product of the g-factor $g$ and the Bohr magneton $\mu_{B}$.
The second term is the Heisenberg Hamiltonian with the magnitude $J$. 
Since the wavenumber $k$ in the anti-crossing region is very small, this term $\sim J(ka)^2$ is neglected henceforth, where $a$ is the lattice constant.
 The third term describes the DDI.
By the Holstein Primakoff transformation\cite{Holstein40} with the annihilation (creation) operator $a_{m}(a_{m}^{\dagger})=S_{mx}+(-)iS_{my}$,
the Hamiltonian (\ref{eq:hamiltonianm}) reduces to~\cite{ClogstonAM56}
\begin{eqnarray}
H_{m}=\sum_{\mathbf{k}}\left[A_{\mathbf{k}}a_{\mathbf{k}}^{\dagger}a_{\mathbf{k}}
+
\frac{1}{2}(B_{\mathbf{k}}a_{\mathbf{k}}a_{-\mathbf{k}}+B_{\mathbf{k}}^{*}a_{\mathbf{k}}^{\dagger}a_{-\mathbf{k}}^{\dagger}) \right],\label{eq:noninteractingH}
\end{eqnarray}
with $A_{\mathbf{k}} 
=A_{0}+B_{\mathbf{k}}$ and $\
B_{\mathbf{k}}
=2\pi \mu M_{s}\sin^2\theta_{\mathbf{k}}\mathrm{e}^{-2i\phi_{k}}$, where $M_{s}$ is the saturation magnetization, $\theta_{\mathbf{k}}=\tan^{-1}\frac{\Tilde{k}}{k_{z}}$, $\Tilde{k}=\sqrt{k_{x}^2+k_{y}^2}$, $k=|\mathbf{k}|$, and $\phi_{k}=\tan^{-1}\frac{k_{y}}{k_{x}}$. $A_{0}=\mu (h+M_{s}N_z)$, $N_{z}$ is the demagnetization coefficient, $S$ is the effective spin ($M_{s}=\frac{N}{V}\mu S$), and $N$ is the number of sites in the volume $V$. 
The eigenvalue is $
E_{m\mathbf{k}}=\sqrt{A_{\mathbf{k}}^2-|B_{\mathbf{k}}|^2}$ by the Bogoliubov transformation,
$
a_{\mathbf{k}}=t_{\mathbf{k}}c_{\mathbf{k}}-s_{-\mathbf{k}}c_{-\mathbf{k}}^\dagger
$
where $c_{\mathbf{k}}$ $ (c_{\mathbf{k}}^\dagger)$ is the bose annihilation (creation) operator, $
t_{\mathbf{k}}=\sqrt{ (A_{\mathbf{k}}+E_{m\mathbf{k}})/2 E_{m\mathbf{k}}} $ and $s_{\mathbf{k}} =(B_{\mathbf{k}}/|B_{\mathbf{k}|})\sqrt{(A_{\mathbf{k}}-E_{m\mathbf{k}})/2 E_{m\mathbf{k}} }
$. 
Due to the external magnetic field $h$, there is a finite gap $E_{m\mathbf{k}=0}$.

\begin{figure}[htbp]
 \begin{center}
 \includegraphics[width=80mm]{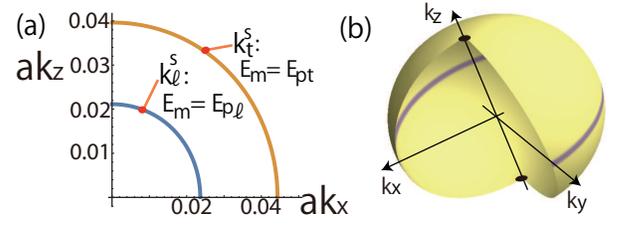}
 \end{center}
 \caption{ 
 (a) Contours of $\mathbf{k}^{s}_{\ell,t}$ of the cross-section of the anti-crossing region given by $E_{m}=E_{p\ell, t}$ at $k_{y}=0$, $k_{x,z}>0$.
 (b) The surface given by $\mathbf{k}^{s}_{\ell}$ for the case with the LA phonon. The contour and dots on the surface show the disappearance of the hybridization $R_{\ell}=0$ on $\mathbf{k}^s_{\ell}$.
Parameters are used for YIG with $h=4\times4\pi M_{s}$ in both of the results (a) (b).
}
 \label{fig:SpinlPhoonFS}
\end{figure}

The dispersion of the phonon is $E_{p\lambda\mathbf{k}}=\hbar c_{p\lambda}k$, where $\lambda$ specifies $\lambda=t_{1,2}$ for TA and $\lambda=\ell$ for LA, and $c_{p\lambda }$ is the velocity.
Due to $E_{m\mathbf{k}=0}\not=0$, crossings of the dispersions of the magnon and phonon occur, and the hybridization is induced.
In Fig.~\ref{fig:SpinlPhoonFS}(a), we show the contours $\mathbf{k}_{\lambda}^{s}$ determined by $E_{m\mathbf{k}_{\lambda}^{s}}=E_{p\lambda\mathbf{k}_{\lambda}^{s}}$ in momentum surface space. We employ $h=4\times4\pi M_{s}$, $N_{z}=0$ by assuming the system to be spherical, and the following parameters appropriate for
yttrium iron garnet (YIG) : $4\pi M_{s}=$0.1750T, $a=12.38${\AA}, the density is $5.17$g/cm$^{3}$, and the velocity of the phonons are 
$
c_{p\ell}=7.209\times10^{5} \mathrm{cm/s}$, $c_{pt}=3.843\times10^{5} \mathrm{cm/s}$ ( $c_{pt }=c_{pt_{1} }=c_{pt_{2} }$).

{\it Interaction between magnons and phonons.--} 
Generally, the exchange coupling $J_{ij}$ depends on the displacement of the atoms, i.e., phonon; namely the exchange striction gives the most ubiquitous magnon-phonon interaction. However, this interaction does not lead to the linear terms in the magnon coordinates, since it preserves the rotational symmetry in the spin space. The SOI~\cite{Kittel58, SchlomannE60} is one of the candidates for breaking the symmetry, and here we also obtain such an interaction by the DDI.
The magnon-phonon coupling through the DDI is given from the Hamiltonian~(\ref{eq:hamiltonianm}) as 
\begin{eqnarray}
H_{\mathrm{dp}}=-\frac{3\sqrt{2S^3}\mu ^2}{2}\sum_{\l\not= m}\frac{z_{lm}
(r_{lm}^{+}a_{l}^{\dagger}+r_{lm}^{-}a_{l})
}{ |\mathbf{r}_{lm}|^5},
\end{eqnarray}
where $r_{lm}^{\pm}=x_{lm}\pm i y_{lm}$.
Expanding $\mathbf{r}_{l m}$ with respect to the phonon coordinates, we obtain 
 \begin{eqnarray}
H_{\mathrm{dp}}
&=&
\sum_{\mathbf{k}\lambda}[
Q_{\lambda \mathbf{k}}c_{\mathbf{k}}^{\dagger}(b_{\lambda\mathbf{k}}+b_{\lambda-\mathbf{k}}^{\dagger})
+c.c.
]\label{eq:DDhamiltonian},
\\
Q_{\lambda \mathbf{k}}&=&-\sum_{i=x,y,z}g_{\mathrm{d} \lambda}\xi_{\lambda i\mathbf{k}}(G_{i\mathbf{k}} t_{\mathbf{k}} 
+G_{i\mathbf{k}}^{*}s_{\mathbf{k}}^{*})\label{eq:Qfunc},
\end{eqnarray}
where $\boldsymbol{\xi}_{\lambda \mathbf{k}}$ is the polarization vector of the phonon and $b_{\lambda \mathbf{k}} (b_{\lambda \mathbf{k}}^{\dagger})$ is the bosonic annihilation (creation) operator of the phonon. By replacing $\mathbf{k}$ with the dimensionless vector as $a\mathbf{k}\to \mathbf{k}$, the coupling constant is given as $g_{\mathrm{d} \lambda}=(\pi \mu M_{s}/2)\sqrt{ \hbar S/aM_{p}c_{p\lambda}
}$, where $M_{p}$ is the mass of the atom.
The functions in Eq.~(\ref{eq:Qfunc}) are given as
$G_{x\mathbf{k}}
=-2i k_z(k_{z}^2-(k_{x}+ik_{y})^2)k^{-\frac{5}{2}}$,
$
G_{y\mathbf{k}}
=2 k_z(k_{z}^2+(k_{x}+ik_{y})^2)k^{-\frac{5}{2}}$, and
$
G_{z\mathbf{k}}
=\frac{4i}{3}k_{z}^2(k_x+ik_y)k^{-\frac{5}{2}}
$. 
By these functions, one can find that hybridization vanishes for $k_{z}=0$ independent of $\boldsymbol{\xi}_{\lambda \mathbf{k}}$.
 Consequently, the anti-crossing does not occur there, and the gapless points form a loop on the $k_{z}=0$ plane.


In real systems, we should take into account both the SOI and DDI for the interaction Hamiltonian as
\begin{eqnarray}
H_{\mathrm{int}}&=&-
\sum_{\mathbf{k}\lambda}[
R_{\lambda \mathbf{k}}c_{\mathbf{k}}^{\dagger}(b_{\lambda\mathbf{k}}+b_{\lambda-\mathbf{k}}^{\dagger})
+c.c.
]
\end{eqnarray}
with $
R_{\lambda\mathbf{k}}=P_{\lambda\mathbf{k}} +Q_{\lambda\mathbf{k}} $, where $P_{\lambda\mathbf{k}}$ is the contribution from the SOI \cite{Kittel58} given as $P_{\lambda\mathbf{k}}=\sum_{i}g_{2 \lambda}\xi_{\lambda i\mathbf{k}}(F_{i\mathbf{k}} t_{\mathbf{k}} 
+F_{i\mathbf{k}}^{*}s_{\mathbf{k}}^{*})$ with $F_{x\mathbf{k}}=-ik_{z}k^{-\frac{1}{2}} $, $F_{y\mathbf{k}}=iF_{x\mathbf{k}}$, and $
F_{z\mathbf{k}}=-i(k_{x}+ik_{y})k^{-\frac{1}{2}}$. The coupling constant is given by the strength of the SOI $b_{2\lambda}$ as 
$g_{2\lambda}=( b_{2\lambda}/ 2) \sqrt{\hbar/S a M_{p}c_{p\lambda
}}$.
In YIG, the ratios of the two coupling constants are given as $\frac{g_{2\ell}}{g_{\mathrm{d}\ell} }=57.3$ and $\frac{g_{2t}}{g_{\mathrm{d}t} }=114.6$, and therefore the SOI is dominant in the interaction process.
We note that the coupling from the DDI is universally 
determined only by the magnitude of the magnetic 
moment and the lattice displacement, while the strength of the SOI 
$b_{2 \lambda}$ depends on the electronic structure 
of each material.
For the LA phonon, the gapless points remain when the two interactions are present. In Fig.~\ref{fig:SpinlPhoonFS}(b), those gapless points are shown on the anti-crossing surface (Fig.~\ref{fig:SpinlPhoonFS}(a)). 
In contrast, for the TA modes, the gapless loop on $k_{z}=0$ disappears due to the SOI, since the hybridization for the $z$-polarization does not vanish, $F_{z\mathbf{k}}\not=0$ for $k_{z}=0$.

We construct an effective model for focusing on the anti-crossing region, where we can neglect the product of two annihilation/creation operators of the magnon and phonon. Since the gap energy is tiny, we can treat each anti-crossing separately. 
The effective Hamiltonians for the LA and TA modes are 
\begin{eqnarray}
H_{\ell}=
\begin{pmatrix}
E_{m\mathbf{k}}& R_{\ell\mathbf{k}}^{*}\\
R_{\ell\mathbf{k}}&E_{p\ell \mathbf{k}}\label{eq:HamiltonianMEl}
\end{pmatrix},\ \ 
H_{t}=
\begin{pmatrix}
E_{m\mathbf{k}}&R_{t_{1}\mathbf{k}}&R_{t_{2}\mathbf{k}}\\
R_{t_{1}\mathbf{k}}^{*}&E_{pt \mathbf{k}}&0\\
R_{t_{2}\mathbf{k}}^{*}&0&E_{pt \mathbf{k}}
\end{pmatrix}.\label{eq:Hamiltonian}
\end{eqnarray}
For eigenstates with the LA mode $|\psi_{\ell \mathbf{k}}^{\pm}\rangle$, the eigenvalues are $ \varepsilon_{\ell \mathbf{k}}^{\pm }=\frac{E_{m \mathbf{k}}+E_{p\ell \mathbf{k}}}{2}\pm\eta _{\ell \mathbf{k}}$. The suffix $+(-)$ corresponds the upper (lower) branch, and $\eta _{\ell \mathbf{k}}=\eta _{\lambda=\ell \mathbf{k}}$ with $\eta _{\lambda \mathbf{k}}=\sqrt{ \delta E_{\lambda \mathbf{k}}^2 +|R_{\lambda \mathbf{k}}|^2 }$ and $\delta E_{\lambda \mathbf{k} }= \frac{E_{m \mathbf{k}}-E_{\lambda \mathbf{k}}}{2}$.
For the TA eigenstates $|\psi_{t0}\rangle$, $|\psi_{t\pm}\rangle$, 
the eigenvalues are $
\varepsilon_{t0 \mathbf{k}}=E_{pt \mathbf{k}}$ and $\varepsilon_{t \mathbf{k}}^{\pm }=\frac{E_{m\mathbf{k}}+E_{pt \mathbf{k}}}{2}
\pm
\eta_{t \mathbf{k}}$ with $|R_{t\mathbf{k}}|^2=|R_{t_{1}\mathbf{k}}|^2+|R_{t_{2}\mathbf{k}}|^2$.
In the following discussion, the Berry curvature is calculated by the polarization vectors given as
$\boldsymbol{\xi} _{\ell\mathbf{k}}=i
\frac{1}{k}(k_{x},k_{y},k_{z})$, $
\boldsymbol{\xi} _{t_{1}\mathbf{k}}=i \frac{1}{ \Tilde{k} } (k_{y},-k_{x},0)$, and $\boldsymbol{\xi} _{t_{2}\mathbf{k}}=\boldsymbol{\xi} _{\ell\mathbf{k}}\times\boldsymbol{\xi} _{t_{1}\mathbf{k}}$.

\begin{figure}[htbp]
 \begin{center}
 \includegraphics[width=85mm]{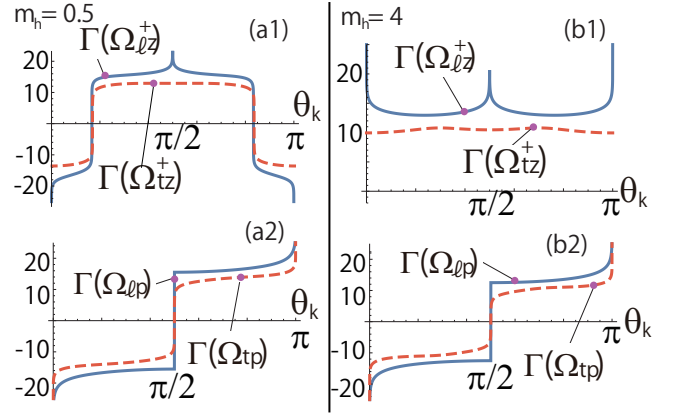}
 \end{center}
 \caption{The $z$-component and in-plane Berry curvature ($\Omega_{\lambda z}^{+}$ and $\Omega_{\lambda p}$) on the anti-crossing region $\theta_{\mathbf{k}}=\tan^{-1} {(\Tilde{k}}/{k_{z})} \in[0,\pi]$ given by $E_{m\mathbf{k}^{s}}=E_{p\lambda\mathbf{k}^{s}}$ for YIG.
These values are shown in a log scale: $\Gamma(\Omega_{\lambda p})$ and $\Gamma(\Omega_{\lambda z}^{+})$ (Eq.~(\ref{eq:defgamma})). 
The solid/dotted curve is the result for the LA/TA mode, and results for $m_{h}\equiv \frac{h}{4\pi M_{s}}=0.5$ and $m_{h}=4$ are shown on the left and right of the vertical line, respectively. 
The sign changes of $\Omega_{\ell z}^{+}$ and $\Omega_{t z}^{+}$ occur at $\theta_{\mathbf{k}}=\frac{\pi}{6},\frac{5\pi}{6}$ (Eq.~(\ref{eq:Rpoints})) for $m_h=0.5$ in (a1), while not for $m_h=4$ in (b1). 
For $\Omega_{\ell z}^{+}$ and $\Omega_{\lambda p}$, the divergences are seen with the gap-closing, while $\Omega_{t p}^{+} $ diverges at $\Tilde{k}=0$ without the gap-closing in (a2) (b2).
}
 \label{fig:BerryYIG}
\end{figure}

{\it Texture of the Berry curvature.--}
By the effective Hamiltonians (\ref{eq:Hamiltonian}), we discuss the Berry curvature $\boldsymbol{\Omega}^{\pm}_{\lambda}$ induced by the hybridization.
By the definition, $\boldsymbol{\Omega}^{\pm}_{\lambda}=i\langle \nabla_{\mathbf{k}}\psi_{\lambda\pm} |\times | \nabla_{\mathbf{k}}\psi_{\lambda\pm} \rangle$~\cite{BerryMV84}, we have
\begin{eqnarray}
\Omega_{\lambda i=x,y}^{\pm}&=&\mp
\frac{k_{i}}{\Tilde{k}}\Omega_{\lambda p}, \ \ 
\Omega_{\lambda z}^{\pm}=\mp\frac{1}{2\Tilde{k}}\partial_{\Tilde{k}} \frac{ \delta E_{\lambda } }{ \eta_{\lambda}} \label{eq:BC},
\end{eqnarray}
where $
\Omega_{\lambda p}\equiv \frac{1}{2\Tilde{k}}\partial_{k_{z}} \frac{ \delta E_{\lambda } }{ \eta_{\lambda}}$ shows the in-plane Berry curvature,
 and $\Omega_{\lambda p}$ and $\Omega_{\lambda z}^{\pm}$ are functions of $(\Tilde{k}, k_{z})$.
We note that $|\psi_{t0}\rangle$ has no Berry curvature for the TA mode, since it does not have the magnon component.
The hybridization $R_{\lambda}$ is much smaller than the dispersions of the magnon and phonons. Therefore, the Berry curvature is enhanced only near the anti-crossing points satisfying $\delta E_{\lambda \mathbf{k}=\mathbf{k}_{\lambda}^{s} }=0$, which describes a surface in momentum space $\mathbf{k}_{\lambda}^{s}$ (Fig.~\ref{fig:SpinlPhoonFS}), and there, the Berry curvature has the form
\begin{eqnarray}
\Omega_{\lambda z}^{\pm}=\mp\frac{1}{2\Tilde{k}} \frac{ \partial_{\Tilde{k}}\delta E_{\lambda } }{ |R_{\lambda}|},
\ 
\Omega_{\lambda p}= \frac{1}{2\Tilde{k}}\frac{\partial_{k_{z}} \delta E_{\lambda } }{ |R_{\lambda}|} .\label{eq:ddE}
\end{eqnarray}
We show the Berry curvature of the upper states $|\psi_{\lambda }^{+}\rangle$, $\Omega_{\lambda p}$ and $\Omega_{\lambda z}^{+}$ along $\mathbf{k}_{\lambda}^{s}$ surface on the $(\Tilde{k}, k_{z})$ plane for $\theta_{\mathbf{k}}=\tan^{-1} \frac{\Tilde{k}^{s}}{k_{z}^{s}} \in[0,\pi]$, in 
Fig.~\ref{fig:BerryYIG} by using parameters of YIG, for $m_{h}\equiv \frac{h}{4\pi M_{s}}=0.5$ (a1)(a2) and $m_{h}=4$ (b1)(b2).
Since those functions rapidly become large close to $\mathbf{k}_{\lambda}^{s}$, the Berry curvature is shown on a log scale by using a function:
\begin{eqnarray}
\Gamma(\Omega)\equiv \mathrm{sign}(\Omega)\log(1+|\Omega|)
\label{eq:defgamma}.
\end{eqnarray}

Next, we analyze the Berry curvature in the vicinity of the gapless points which are referred to as the Weyl points in fermionic systems~\cite{BurkovAAl11}.
For the LA mode, the Berry curvature diverges on the $k_{z}$ and $k_{x}$ axes ($\theta_{\mathbf{k}}=0,\frac{\pi}{2}, \pi$) where the energy difference $\delta E_{\ell\mathbf{k} }$ and the hybridization $R_{\ell\mathbf{k}}$ simultaneously vanish, i.e., $\delta E_{\ell}=R_{\ell}=0$.
 We calculate the Berry curvature in the vicinity of those singular points and find that they are highly anisotropic as described below, unlike typical singular points such as Weyl nodes of electronic systems~\cite{BurkovAAl11}.
Both $\delta E_{\ell }$ and $R_{\ell} $ vanish simultaneously at~\cite{AndaE76}
\begin{eqnarray}
\pm \mathbf{k}_{\mathrm{D}}=\left(0,0,\pm k_{\mathrm{D}} \right),
\mathbf{k}_{\mathrm{L}}(\phi_{k})\equiv \Tilde{k}_{\mathrm{L}}\left(\cos\phi_{k},\sin\phi_{k},0 \right),
\end{eqnarray}
 where $k_{\mathrm{D}}\equiv \frac{\mu h}{\hbar c_{p\ell}}$,
 $\Tilde{k}_{\mathrm{L} }\equiv \frac{\sqrt{\mu h(\mu h+4\pi \mu M_{s})}}{\hbar c_{p\ell}}$, and $^\forall \phi_{k} \in[0,2\pi]$.
 The gapless points given by $\mathbf{k}_\mathrm{L}(\phi_{k})$ form a loop on the plane $k_{z}=0$ (Fig.~\ref{fig:SpinlPhoonFS}(b)).
 Close to $\pm \mathbf{k}_{\mathrm{D}}$, the effective Hamiltonian further reduces to the Weyl Hamiltonian with opposite Weyl charges~\cite{BurkovAAl11} on $+\mathbf{k}_{\mathrm{D}}$ and $-\mathbf{k}_{\mathrm{D}}$.
For $\mathbf{k}\sim+\mathbf{k}_{\mathrm{D}}$, the Berry curvature has the form
\begin{eqnarray} 
\Omega_{\ell z }^{\pm}
=\mp 
\frac{g_{\mathrm{D}}^2}{2}\frac{ \cos\theta_{\mathrm{D} } }{\eta_{\ell} ^2 },
 \ 
\Omega_{\ell p}
=-\frac{\hbar c_{p\ell} g_{\mathrm{D}}}{4} \frac{ \sin\theta_{\mathrm{D}} }{\eta_{\ell} ^2 }
\label{eq:BCdot},
\end{eqnarray}
where $\eta_{\ell}=\sqrt{\left(\hbar c_{p\ell}/2 \right)^2\delta k_{z}^2 +g_{\mathrm{D}}^2\Tilde{k}^2 }$ with $g_{\mathrm{D}} =\frac{2 (g_{\mathrm{d}\ell }-g_{2\ell })}{\sqrt{k_{\mathrm{D}}}}$, $\theta_{\mathrm{D}}=\tan^{-1}\frac{2 g_{\mathrm{D}}\delta \Tilde{k}}{\hbar c_{p\ell} \delta k_{z}}$, and $(\delta \Tilde{k}, \delta k_{z})$ is the infinitesimal wavevector from the singular points.
The Berry curvature is given by $(\Omega_{\ell p}, 
\Omega_{\ell z }^{\pm})=\left(0,\mp2g_{D}^2/(\hbar c_{p\ell} \delta k_{z})^2\right)$ for  $\Tilde{k}=0$ and $(\Omega_{\ell p},
\Omega_{\ell z }^{\pm})=(-(\hbar c_{p\ell}/g_{D})/ \Tilde{k}^2,0)$ for $\delta k_{z}=0$, respectively.
Although the Berry curvature is represented as the hedgehog-like texture of the order of $\delta k^{-2}(\sim\eta_{\ell}^{-2})$, the direction of the Berry curvature is almost perpendicular to the $z$-direction, because the hybridization is much smaller
 than the energy scale of the phonon, $\hbar c_{p\ell} \gg g_{\mathrm{D}} $.
Around $\mathbf{k}_\mathrm{L}(\phi_{k})$,
 we have 
\begin{eqnarray}
\Omega_{\ell z}^{\pm}= \mp 
\frac{\hbar c_{p\ell}}{4\Tilde{k}_{\mathrm{L}}}\frac{ \cos^2\theta_{\mathrm{L} } }{\eta_{\ell} }
,
 \ 
\Omega_{\ell p}
=\frac{g_{\mathrm{L}}}{2\Tilde{k}_{\mathrm{L}}} \frac{ \cos\theta_{\mathrm{L} }\sin\theta_{\mathrm{L} } }{\eta_{\ell} },\label{eq:BCloop}
\end{eqnarray}
where $\eta_{\ell}=\sqrt{\left(\hbar c_{p\ell}/2 \right)^2\delta \Tilde{k}^2 +g_{\mathrm{L}}^2k_{z}^2 }$, $g_{\mathrm{L}}
=g_{\mathrm{D}}\frac{k_{\mathrm{D}}}{\Tilde{k}_{\mathrm{L} }}$, and $\theta_{\mathrm{L}}=\tan^{-1}\frac{\hbar c_{p\ell} \delta \Tilde{k}}{2g_{\mathrm{L}}\delta k_{z}}$. Due to $\hbar c_{p\ell}\gg g_{\mathrm{L}} $, the direction of the Berry curvature vector is along the $z$ axis, similar to the previous case.
Therefore, the two singular points are different in the behaviors of the divergence.

For the case of the TA mode, there are neither Weyl nodes nor singular loops, except for special cases of $g_{2t}=0$ or $g_{2t}=-2g_{\mathrm{d}t}$.
Nonetheless, the divergence occurs for $\Omega_{t p}$ on the $k_{z}$ axis as shown in Figs.~\ref{fig:BerryYIG}(a2, b2). Close to $\Tilde{k}=0$, it is described by Eq.~(\ref{eq:ddE}) as $
\Omega_{tp}= -\frac{1}{2\Tilde{k}}\frac{k_{z}}{|k_{z}|}\chi_{k_{z}},
$
 with $\chi_{k_{z}}=\frac{(g_{2t }+2g_{\mathrm{d}t })^2(\mu h+\hbar c_{pt} |k_{z}|)/2}{
\sqrt{(\mu h-\hbar c_{pt}|k_{z}|)^2/4+2g_{t}^2|k_{z}|}}$ converging on $\Tilde{k}=0$. Then, $\Omega_{tp}$ is on the order of $\delta k^{-1}$.
Meanwhile, $\Omega_{tz}^{\pm}$ does not diverges at $\Tilde{k}=0$ (Fig.~\ref{fig:BerryYIG}(a1, b1)), and the Berry curvature is distributed almost in the radial direction perpendicular to the $k_{z}$ axis.
This divergence of the Berry curvature is unconventional since the gap remains open and is attributed to the singular DDI along the $k_{z}$ axis.

{\it Observation and magnon current control.--}
We give an observation scheme of the Hall effect of the quasiparticle consisting of the magnon and phonons. 
When the applied magnetic field $h$ along the $z$-direction has a spatially weak gradient, the magnon feels the force $\mathbf{F}=-\mu \nabla h(\mathbf{r})$, and the anomalous velocity $
\mathbf{v}_{A\lambda}^{\pm} =\frac{\mathbf{F}}{\hbar}\times\boldsymbol{\Omega}^{\pm}_{\lambda}
$ is induced~\cite{XiaoD10}.
 The impulsive Brillouin scattering can excite the phonons and magnons through hybridization with momentum resolution~\cite{DemokritovSO01,SergaAA10,YanXY87}, for example.
In actual experiment, the phonon is excited as a wavepacket with a finite distribution corresponding to the momentum resolution, and the Hall current emerges as the integral of Berry curvature. We consider the case where the wavepacket has the width $w$ in momentum space and the 
 center of the momentum is near the anti-crossing point: $\mathbf{k}_{\lambda=\ell,t}^{s}$ ($E_{m\mathbf{k}_{\lambda}^{s}}=E_{p\lambda\mathbf{k}_{\lambda}^{s}}$).
On $\mathbf{k}_{\lambda}^{s}$, the Berry curvature by Eq.~(\ref{eq:ddE}) is concentrated in the momentum region of the width $w_{\lambda}\sim\frac{|R_{\lambda\mathbf{k}_{\lambda}^{s}}|}{
\sqrt{(\partial_{k_{z}}\delta E_{\lambda})^2+(\partial_{\Tilde{k}}\delta E_{\lambda})^2}
}$. 
For $w<w_{\lambda}$, we can observe large anomalous velocities corresponding to the distribution of the Berry curvature along the anti-crossing region. 
For $w\gg w_{\lambda}$, the averaged anomalous velocity over the wavepacket is estimated as $|\langle\mathbf{v}_{A\lambda \mathbf{k}}\rangle|\sim \left|\frac{\mathbf{F}}{\hbar}\times\boldsymbol {\Omega}^{\pm}_{\lambda \mathbf{k}_{\lambda}^{s}}\frac{w_{\lambda}}{w}\right|=\frac{|\mathbf{F}|}{2\hbar \Tilde{k}_{\lambda}^{s}w} $ for $\Tilde{k}_{\lambda}^{s}>w$.
For both the modes, the averaged Berry curvature vanishes at $\Tilde{k}_{\lambda}^{s}=0$, because the cancellation occurs when the width includes the singular point itself, and the maximum anomalous velocity is observed when the distance from the center to the $k_{z}$ axis is the width of the wavepacket; namely $\Tilde k_{\lambda}^{s}$ in the above estimation should be replaced with $w$ as $|\langle\mathbf{v}_{A\lambda \mathbf{k}}\rangle|\sim \frac{|\mathbf{F}|}{2\hbar w^2}$.
For $w\gg w_{\lambda}$, the excitation involves states on the upper and lower branches of the anti-crossing.
In this case, the two excited waves have opposite directions of the anomalous velocity, and then they will be observed by detectors spatially far from the excitation point.

For the strong SOI $g_{2\lambda}>g_{\mathrm{d}\lambda}$, in the strong magnetic field $m_{h}>1$, the following approximation is valid: $w_{\lambda} \sim\frac{b_{\lambda}}{\hbar c_{p\lambda}^2}\sqrt{\frac{\mu h}{SM_{p}}}$, $k_{\lambda}^{s}\sim\frac{\mu h}{\hbar c_{p\lambda}}$.
By the approximation, we obtain the group velocity of the magnon $|v_{m\lambda}| = \frac{4\pi M_{s}}{h}c_{p\lambda}|\cos\theta_{k}|\sin\theta_{k}$ which depends on $\theta_{\mathbf{k}}$ due to the DDI; $|v_{m\lambda}|$ vanishes at $\Tilde{k}=0$ or $k_{z}=0$. 
Therefore, to obtain the large Hall angle, small values of $\Tilde{k}^{s}$ are favorable, and due to $c_{p\ell}>c_{pt}$, the magnon interacting with the TA phonon is are favorable for the observation of the Hall current. For $\Tilde{k}_{\lambda}^{s}=w (\ll k_{\lambda}^{s})$ where the maximum anomalous velocity is obtained, the group velocity is obtained as $|v_{m\lambda}|\sim\frac{4\pi\hbar \mu M_{s}c_{p\lambda}^2 w}{ (\mu h)^2} $.
Now, we consider a setting: the momentum resolution $w= 10^4$cm$^{-1}$, and the gradient of the magnetic field $|\nabla h|=1$T/cm~\cite{HiroseR04, CoeyJMD02} for YIG in $h=1$T.
Then, the estimations are the following: the separation between the two surfaces $|\mathbf{k}_{t}^{s}-\mathbf{k}_{\ell}^{s}|\sim10^{5}$cm$^{-1}$, $w_{\ell}, w_{t}\sim10^2, 10^3$cm$^{-1}$, $|v_{m\ell}| ,|v_{mt}| \sim 10^4, 10^3$cm/s.
The maximum anomalous velocity is estimated of the order of $|\langle\mathbf{v}_{A\lambda \mathbf{k}}\rangle| \sim \frac{|\mathbf{F}|}{2\hbar w^2}\sim 10^3$cm/s. 
$|\langle\mathbf{v}_{A\lambda \mathbf{k}}\rangle|$ is comparable to the group velocity, and the the Hall signals will clearly be observed, especially for the case with TA mode.

Finally, we discuss the rapid sign change of the Berry curvature.
The in-plane Berry curvature $\Omega_{\lambda p}$ changes its sign on $k_{z}=0$ as shown in Figs.~\ref{fig:BerryYIG}(a2, b2). $\Omega_{\lambda p}$ cannot avoid this sign change in momentum space due to the relation $\Omega_{\lambda p(\Tilde{k},-k_{z})}=-\Omega_{\lambda p(\Tilde{k},k_{z})}$.
On the other hand, for the $z$-component Berry curvature $\Omega_{\lambda z}^{\pm}$, the emergence of the sign change depends on the ratio $m_{h}=\frac{h}{4\pi M_{s}}$;
 one can see the sign changes in Fig.~\ref{fig:BerryYIG}(a1) while not in (b1).
By Eq.~(\ref{eq:ddE}), the sign change on the two momenta $(\Tilde{k},k_{z})=(\Tilde{k}^{r}_{\lambda}, +k_{\lambda z}^{r})$ and $(\Tilde{k}^{r}_{\lambda}, - k_{\lambda z}^{r})$, which are given by
\begin{eqnarray}
\Tilde{k}^{r}_{\lambda}=\frac{\mu h }{2\hbar c_{p\lambda}}  \sqrt{\frac{1-m_{h}^2}{m_{h}} },\ 
k_{\lambda z}^{r}=\frac{\mu h }{2\hbar c_{p\lambda}} \frac{1+m_{h}}{\sqrt{m_{h}}}.
\label{eq:Rpoints}
\end{eqnarray}
The above equations are valid only when $m_{h}<1$, and this is the condition for the sign change of $\Omega_{\lambda z}^{\pm}$.
For  $m_{h}=0.5$, those points appear at $\theta_{(\Tilde{k}^{r}_{\lambda}, k_{\lambda z}^{r})} =\tan^{-1}(\Tilde{k}^{r}_{\lambda}/k_{\lambda z}^{r})=\frac{\pi}{6}$ and $\theta_{(\Tilde{k}^{r}_{\lambda}, -k_{\lambda z}^{r})} =\frac{5 \pi}{6}$, as shown in Fig.~\ref{fig:BerryYIG}(a1).
The sign change
 occurs rapidly on the anti-crossing region because the value of the Berry curvature is inversely proportional to the hybridization $|R_{\lambda}|^{-1}$ (Eq(\ref{eq:ddE})). In particular, for $\Omega_{\ell p}$, the sign change more drastically occurs due to $|R_{\ell}|=0$ at $k_{z}=0$.
These sign inversions give rise to the abrupt change of the Hall current in opposite direction, when the wavenumber of the wavepacket is gradually varied along the anti-crossing surface $\mathbf{k}_{\lambda}^{s}$.
For observing the switching of the Hall signal, the following conditions are required for the momentum resolution $w$.
For the Hall current by $\Omega_{\lambda p}$,
$w$ should be smaller than the width of the anti-crossing region along the $z$-direction: namely $w<\mu h/\hbar c_{p\lambda}$ at $m_{h}\geq 1$, and $w<k_{\lambda z}^{r}$ at $m_{h}< 1$ (Eq.(\ref{eq:Rpoints})).
For $\Omega_{\lambda z}$, the condition for the resolution is given as $w< \Tilde{k}^{r}_{\lambda}$,
which is estimated on the order of $\Tilde{k}^{r}_{t,\ell}\sim 10^4$cm$^{-1}$ for $m_{h}=\frac{1}{\sqrt{3}}$. 
The switching of the Hall signal has potential for various applications of magnonics.


{\it Acknowledgments.--} 
We thank N. Ogawa for useful discussions.
RT was supported by Grant-in-Aid for Japan
Society for the Promotion of Science Fellows No. 25-9798.
This work was supported by JSPS KAKENHI (Grant No. 26103006) and Grant-in-Aids for Scientific Research (S) (No.~24224009) 
from the Ministry of Education, Culture, Sports, Science and Technology (MEXT) of Japan.

%
%


\end{document}